\documentclass[journal]{IEEEtran}

\usepackage{amsmath}

\usepackage{url}
\usepackage{graphicx}  
\usepackage{float}  
\usepackage{enumitem} 

\begin{document}

\title{Improving Discovery of Known Software Vulnerability For Enhanced Cybersecurity}

\author{Devesh Sawant, Manjesh K. Hanawal, and Atul Kabra\\
        MLiONS Lab, Department of IEOR, IIT Bombay, Mumbai\\
}


\maketitle
\begin{abstract}

Software vulnerabilities are commonly exploited as attack vectors in cyberattacks. Hence, it is crucial to identify vulnerable software configurations early to apply preventive measures. Effective vulnerability detection relies on identifying software vulnerabilities through standardized identifiers such as Common Platform Enumeration (CPE) strings. However, non-standardized CPE strings issued by software vendors create a significant challenge. Inconsistent formats, naming conventions, and versioning practices lead to mismatches when querying databases like the National Vulnerability Database (NVD), hindering accurate vulnerability detection.

Failure to properly identify and prioritize vulnerable software complicates the patching process and causes delays in updating the vulnerable software, thereby giving attackers a window of opportunity. To address this, we present a method to enhance CPE string consistency by implementing a multi-layered sanitization process combined with a fuzzy matching algorithm on data collected using Osquery. Our method includes a union query with priority weighting, which assigns relevance to various attribute combinations, followed by a fuzzy matching process with threshold-based similarity scoring, yielding higher confidence in accurate matches. Comparative analysis with open-source tools such as FleetDM demonstrates that our approach improves detection accuracy by 40\%.

\end{abstract}


\section{Introduction}

\IEEEPARstart{S}{oftware} vulnerabilities are a pervasive issue in the modern digital landscape, posing significant security risks to organizations across industries. Vulnerabilities are typically defined as weaknesses or flaws in software code that, if exploited by an attacker, can compromise the confidentiality, integrity, and availability of data. Exploitable vulnerabilities are highly sought after by malicious actors aiming to infiltrate systems, steal sensitive data, disrupt operations, or install malware. High-profile incidents of cyberattacks exploiting known vulnerabilities, such as Heartbleed~\cite{heartbleed} and Log4j~\cite{log4j}, demonstrate the necessity of a robust vulnerability management process to safeguard systems and reduce the risk of breaches.

Vulnerabilities are frequently leveraged in attacks mapped to MITRE ATT\&CK~\cite{mitre_attack} TTPs. For instance, tactics such as \textit{Initial Access (TA0001)}, \textit{Execution (TA0002)}, and \textit{Privilege Escalation (TA0004)} often exploit unpatched vulnerabilities. Techniques like \textit{Exploitation of Remote Services (T1210)}, \textit{Valid Accounts (T1078)}, and \textit{Vulnerability Scanning (T1595.002)} highlight the critical role of vulnerabilities in the early stages of cyberattacks. By mapping these techniques, organizations can better understand how vulnerabilities serve as key enablers for sophisticated threat actors.

Organizations that depend on complex, interconnected software systems are particularly vulnerable, as a single unpatched vulnerability can serve as an entry point for widespread compromise. Consequently, efficient detection of vulnerabilities is crucial for risk management, allowing security teams to address weaknesses before they are exploited. Vulnerability detection tools frequently use standardized identifiers known as \textit{Common Platform Enumeration} (CPE)~\cite{cpe_specification} strings to map software and hardware to known vulnerabilities cataloged in resources such as the \textit{National Vulnerability Database} (NVD)~\cite{nvd}. Each vulnerability in the NVD is associated with a \textit{Common Vulnerability and Exposure} (CVE) identifier, which links it to affected software configurations through CPE strings.


\subsection{Motivation and Problem Statement}

The motivation behind this study is rooted in the gap between the idea of CPE string standardization and the reality of software vendor’s inconsistent implementations. Many vendors follow proprietary naming and versioning practices, leading to significant variations in datasets used for vulnerability detection. This lack of consistency becomes a barrier to effective vulnerability management, as security teams may miss vulnerabilities when exact CPE string matches are required. Open-source platforms, including FleetDM, rely on exact or near-exact string matching methods. While FleetDM is effective for software monitoring, it lacks advanced mechanisms to accommodate CPE string variations, which can result in missed detections and potential risks for organizations.

\subsection{Objectives and Contributions}

In this work, we aim to improve the detection of software vulnerabilities by addressing the challenges posed by non-standardized CPE strings. Our primary contributions include:
\begin{itemize}
\item We develop a scalable system architecture that integrates data collection, sanitization, query processing, and fuzzy matching into a seamless pipeline for real-time vulnerability management.
\item We perform a multi-step sanitization process to standardize software names, versions, and vendor information, thereby addressing common discrepancies that arise in real-world datasets
    \item  To retrieve a list of potential matches, we employ a union query that assigns weights based on the parameters matched (e.g., title, vendor, version). These weights prioritize entries and help establish thresholds for subsequent fuzzy matching, enhancing detection accuracy by focusing on highly relevant candidates.

\item We develop a fuzzy matching algorithm capable of tolerating minor variations in naming conventions and develop a scoring mechanism that indicates confidence in the matching outcomes.
    
\item We demonstrate improvement in vulnerability detection rates over standard methods, using FleetDM~\cite{fleetdm_guide} as a benchmark for comparison.
\end{itemize}

Our approach significantly enhances the accuracy of vulnerability detection and provides a proactive tool for organizations to manage software risks. 

\section{Related Work}
Accurate discovery software with known vulnerability relies on the precise identification of software components using standardized identifiers, such as CPE strings. Existing solutions, including FleetDM ~\cite{fleetdm_guide}, utilize basic string matching and sanitization techniques to link software products with vulnerabilities in databases such as NVD. However, these methods often fall short due to variations in software names, versions, and vendor details that lead to mismatches. In the following, we review approaches used in the open-source project FleetDM, its limitations, and the motivations for our enhanced methodology. The primary reason to focus on FleetDM is that full details are available, and we could reproduce the outcomes for a fair comparison. Though several proprietary solutions are available for software vulnerability detection, such as Tenable Nessus~\cite{tenable_io} and CrowdStrike Falcon Spotlight \cite{crowdstrike_falcon}, they are proprietary, and full details about their mechanism are not available. 

\subsection{FleetDM’s Approach to Vulnerability Detection}
FleetDM~\cite{fleetdm_guide} is an open-source endpoint security platform that provides an elementary solution to software vulnerability detection by matching software installed on client systems against known vulnerabilities in the NVD. FleetDM’s approach focuses on basic sanitization techniques to normalize software names, vendor information, and versions to improve the accuracy of database lookups.

\subsubsection{Basic Sanitization in FleetDM}
FleetDM's method centers on standardizing software names by removing common suffixes, such as "Inc." or "LLC," and normalizing cases to lowercase. It also simplifies version numbers by trimming trailing zeros, while removing platform-specific tags and unnecessary metadata where possible. For example:
\begin{itemize}
    \item The software name "OpenVPN Technologies, Inc." is reduced to "openvpn" by removing extraneous terms and standardizing capitalization.
\end{itemize}

\subsubsection{SQL Union-Based Query Matching}
FleetDM employs union-based SQL queries to match software records against NVD data. These queries rely on different combinations of attributes, such as software name, vendor, and version, with priority weighting to increase matching accuracy. For instance:
\begin{itemize}
    \item Priority is given to queries matching the \texttt{title}, \texttt{vendor}, and \texttt{version} fields.
    \item Lower-weighted queries might match only the software name and version, providing a fallback if the full title or vendor information does not match.
\end{itemize}

\subsection{Limitations of FleetDM’s Approach}
While FleetDM's methodology provides a functional baseline, it encounters limitations when handling non-standardized CPE strings or complex versioning schemes. These limitations hinder its ability to accurately detect vulnerabilities in cases where software vendors use inconsistent or abbreviated naming conventions. The main limitations include:

\subsubsection{Insufficient Sanitization and Normalization}
FleetDM’s basic sanitization is effective for handling straightforward discrepancies, such as minor spelling variations or the removal of suffixes. However, it does not fully address the range of inconsistencies encountered in real-world data. For instance:
\begin{itemize}
    \item Software titles that differ significantly due to rebranding or regional naming may not match. For example, "Mozilla Thunderbird" and "Mozilla" might be treated as distinct entries, even though they refer to the same product.
    \item Platform-specific tags or version indicators, such as "beta" or "RC1," may remain after sanitization, resulting in mismatches with standardized database entries.
\end{itemize}

\subsubsection{Lack of Fuzzy Matching}
FleetDM’s approach relies exclusively on exact matches, even when using union queries with flexible attribute combinations. This method fails to account for minor discrepancies in software names, vendor details, or version formats, leading to missed vulnerabilities. In cases where slight variations exist, FleetDM may overlook potential vulnerabilities, as illustrated below:
\begin{itemize}
    \item Small typographical differences, such as "Open VPN" versus "OpenVPN," may cause records to be overlooked.
    \item Variants like "Microsoft Corporation" and "Microsoft Corp." are treated as separate entities without fuzzy matching, reducing the system’s ability to detect vulnerabilities accurately.
\end{itemize}

\subsection{Need for an Enhanced Approach}
Given these limitations, there is a pressing need for a methodology that combines comprehensive sanitization with fuzzy matching techniques to improve detection accuracy. Addressing variations in software titles, vendor names, and versions requires a solution that can adapt to inconsistencies and accurately match software products with corresponding vulnerabilities. Our proposed solution builds upon FleetDM’s foundational structure but introduces a more rigorous sanitization process and fuzzy matching to address data discrepancies.

In the following sections, we present an enhanced methodology that integrates these advanced techniques, resulting in significantly improved detection rates and reduced false negatives.
\section{Proposed Methodology}

The proposed methodology addresses the limitations of existing systems by introducing a structured approach that includes data collection, sanitization, prioritized union queries, and fuzzy matching. These steps work unitedly to improve the accuracy of vulnerability detection and ensure a consistent matching process. This section explains each component in detail.

\subsection{Data Collection}
The data collection process begins with the use of \texttt{osquery}~\cite{osquery_doc}, an open-source tool designed to gather system information from client devices. \texttt{osquery} queries are used to collect data on installed software, including attributes such as software name, version, and vendor, which are then stored in a \texttt{programs} table. This table serves as the primary source of input data for identifying and matching software products against vulnerabilities.

By using \texttt{osquery}, we ensure lightweight and efficient data collection that has minimal impact on system performance, even when deployed across a large number of client systems~\cite{vajra_edr}.

\subsection{Sanitization Process}

The sanitization process is essential for normalizing the collected data and addressing inconsistencies in software names, versions, and vendor information, which frequently arise from variations in how vendors label their products. This process consists of the following steps:

\begin{itemize}
    \item \textbf{Standardizing Software Names}: Extraneous terms such as "Technologies," "Inc.," and "LLC" are removed from software names. Additionally, any architecture strings (e.g., "x86"), language codes, general remarks (e.g., "7-zip - The best software"), and suffixes like ".app" are eliminated. Parentheses and their contents are removed, extra spaces are trimmed, and names are lowercased. Special characters and spacing issues are also corrected, resulting in a consistent format. For example, "OpenVPN Technologies, Inc." becomes simply "openvpn."

    \item \textbf{Normalizing Vendor Information}: Vendors may be recorded under multiple variations, leading to matching errors. The sanitization process normalizes vendor names, transforming differences (e.g., "Microsoft Corp." and "Microsoft Corporation") into a uniform format, such as "Microsoft."

    \item \textbf{Simplifying Version Numbers}: Version numbers can contain unnecessary tags like beta indicators, build numbers, platform identifiers, and other details. The sanitization process trims these to focus on the core version. For instance, "2.4.11-I602-Win10" is simplified to "2.4.11," ensuring a more accurate match with database entries.
\end{itemize}

This comprehensive sanitization process prepares the data for the query stage, significantly reducing discrepancies that could otherwise prevent accurate matching with records in the National Vulnerability Database (NVD).

\subsection{Union Query and Priority Weighting}
The system employs SQL \texttt{UNION} queries in the CPE (Common Platform Enumeration) database to retrieve potential matches, with each query structured to prioritize specific attributes based on their significance to accurate matching. Each query tier is assigned a weight, with lower weights reflecting higher confidence in the match. This weighted approach ensures that entries most likely to match the software details are prioritized, improving detection accuracy and reducing false positives.

The query structure is organized as follows:
\begin{itemize}
    \item \textbf{Weight 1 (High Priority)}: This query matches entries based on the combination of title, vendor, and version, generally providing the most accurate results and highest confidence.
    \item \textbf{Weight 2 (Moderate Priority)}: Matches based on product, vendor, and version, accommodating cases where product names may vary slightly or lack uniformity across sources.
    \item \textbf{Weight 3 (Low Priority)}: Considers both title, product, and version fields with relaxed matching criteria, capturing entries that meet most but not all exact fields.
    \item \textbf{Weight 4 (Fallback Priority)}: Matches on title and version only, which serves as a fallback when vendor data is unavailable or vendor inconsistencies prevent a direct match.
\end{itemize}

This weighted query setup enables a comprehensive yet prioritized list of candidate CPEs for each software entry. By ordering results according to confidence, the system effectively filters out less relevant entries and prioritizes highly probable matches, supporting robust and precise vulnerability detection.

\subsection{Fuzzy Matching with RapidFuzz}
Following the prioritized union query, a fuzzy matching process refines the list by comparing retrieved CPE entries to software names from the programs table. This stage leverages the \texttt{RapidFuzz}~\cite{rapidfuzz} library's \texttt{fuzz::ratio()} function to compute similarity scores between the sanitized software name and corresponding fields (title or product) in the CPE database.

The fuzzy matching workflow proceeds as follows:
\begin{itemize}
    \item \textbf{Similarity Calculation}: For each candidate CPE entry, fuzzy matching calculates a similarity score between the sanitized software name and the retrieved title. When applicable, it also compares the sanitized name to the product field, increasing the robustness of the matching process.
    \item \textbf{Threshold-Based Filtering}: Candidate entries are filtered according to predefined similarity thresholds based on priority weights. For high-confidence matches (weight 1), the threshold is set to 70\% or higher to ensure accuracy. Lower weights employ slightly reduced thresholds, accommodating cases with less detailed information.
    \item \textbf{Selection of Optimal Match}: The system records the highest similarity score across all results. The CPE entry that achieves the top similarity score and meets the threshold is selected as the most likely match, ensuring a high degree of accuracy in the final CPE-to-CVE mapping.
\end{itemize}

By combining weighted union queries with fuzzy matching, the system significantly enhances its ability to detect vulnerabilities accurately, even when minor inconsistencies in naming or formatting are present. This integrated approach supports precise and effective vulnerability detection, directly addressing the challenges associated with non-standardized CPE strings and providing a substantial improvement over traditional exact-match methods.

\subsection{CPE to CVE Mapping}
Once the \texttt{cpe\_string} is stored in the \texttt{programs} table, the next step is mapping each \texttt{cpe\_string} to the National Vulnerability Database (NVD) to retrieve relevant Common Vulnerabilities and Exposures (CVE) identifiers. This mapping process ensures that each software component is linked to current vulnerability information, facilitating accurate and timely threat detection.

The mapping process includes the following steps:

\begin{itemize}
    \item \textbf{Utilizing a Mapping Engine}: The mapping engine takes each \texttt{cpe\_string} and queries the NVD to retrieve any corresponding CVE entries. This automated retrieval ensures that the system maintains up-to-date vulnerability data for each software component.

    \item \textbf{Storing CVE Identifiers}: Once CVE identifiers are retrieved, they are stored in the system's database and associated with the respective software in the \texttt{programs} table. This structured storage allows for quick access to vulnerability details linked to each application.

    \item \textbf{Dashboard Presentation}: The identified vulnerabilities, now mapped to their associated software, are displayed on a user-accessible dashboard. This dashboard provides security teams with a clear view of vulnerabilities, allowing for prioritization and analysis of mitigation steps.
\end{itemize}

This CPE-to-CVE mapping process enhances the system’s detection capabilities, enabling security teams to quickly access actionable vulnerability information for more effective threat response.
\begin{figure}
    \centering
    \includegraphics[height=9cm, width=9cm]{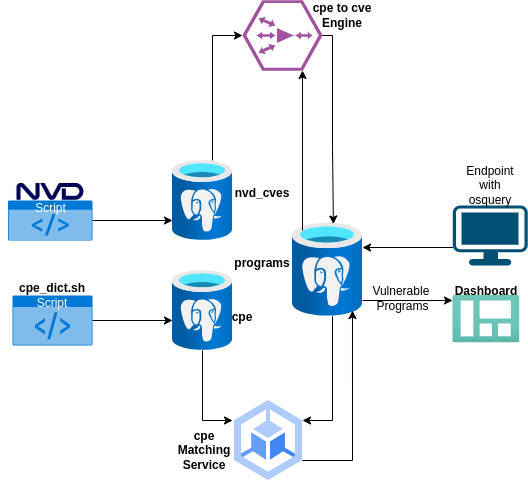}
    \caption{System Architecture for CPE to CVE Detection}
    \label{fig:system_architecture}
\end{figure}

\subsection{System Architecture}

The system architecture for linking Common Platform Enumeration (CPE) identifiers with Common Vulnerabilities and Exposures (CVE) records is outlined as follows:

\begin{enumerate}
    \item \textbf{Data Collection Scripts:} 
    \begin{itemize}
        \item We use Osquery~\cite{osquery_doc} for efficient installed software data collection from the endpoints.
        \item \textbf{NVD Script:} Fetches CVE information from the National Vulnerability Database (NVD) and populates the \texttt{nvd\_cves} table with details like CVE IDs, descriptions, and severity levels.
        \item \textbf{CPE Dictionary Script (cpe\_dict.sh):} Retrieves the CPE dictionary, which standardizes identifiers for software and hardware platforms. This data is stored in the \texttt{cpe} table, mapping software programs to CPEs.
    \end{itemize}

    \item \textbf{Database (PostgreSQL):}
    \begin{itemize}
        \item The database contains three primary tables:
        \item \texttt{nvd\_cves}: Stores vulnerability data from NVD, detailing CVE information.
        \item \texttt{cpe}: Holds CPE records, linking software identifiers to their respective CPE strings.
        \item \texttt{programs}: Contains data on software installed across monitored systems, including software name, vendor, and version.
    \end{itemize}

    \item \textbf{CPE Matching Service:}
    \begin{itemize}
        \item Processes data in the \texttt{programs} table by sanitizing software names, vendors, and versions to standardize the format.
        \item Utilizes a fuzzy matching algorithm to find potential CPE matches, ensuring accurate mappings despite inconsistencies in software names, vendors, or versions.
    \end{itemize}

    \item \textbf{CPE to CVE Mapping Engine:}
    \begin{itemize}
        \item Maps identified CPEs to known CVEs from the \texttt{nvd\_cves} table by cross-referencing data, producing a list of vulnerable programs based on the latest NVD information.
        \item This process indicates which installed software has known vulnerabilities, based on the CPE matching and provides the associated  CVE data.
    \end{itemize}

    \item \textbf{Dashboard:}
    \begin{itemize}
        \item Displays the list of vulnerable programs identified in the system, providing security teams with insights into each program’s associated CVEs and severity.
    \end{itemize}
\end{enumerate}

Overall, this architecture streamlines vulnerability detection by collecting, sanitizing, matching, and mapping software data to vulnerabilities, making critical information readily available for security response.
\section{Results and Analysis}

In this section, we present the results of the vulnerability detection capabilities of two systems: FleetDM, an open-source endpoint security tool, and our custom solution, Vajra, which incorporates enhanced sanitization and fuzzy matching algorithms. We analyze the performance based on sample datasets, calculate detection rates for both systems, and assess the impact of Vajra’s additional sanitization and fuzzy matching steps on overall accuracy.

\subsection{Application Sample Spaces}

We conducted tests on two sample datasets of software applications to evaluate the detection accuracy of FleetDM and Vajra. Each dataset included a mix of commonly used applications with known vulnerabilities, as well as less common software versions that might not be directly listed in the National Vulnerability Database (NVD) or might require extensive sanitization and matching to detect.
\begin{table}[H]
\centering
\caption{Application Sample Space 1}
\small 
\begin{tabular}{|l|c|c|c|}
\hline
\textbf{Application} & \textbf{Version} & \textbf{FleetDM}& \textbf{Vajra}\\
\hline
Adobe Acrobat (64-bit) & 5.0                  & No  & Yes  \\
VLC Media Player       & 1.0.3                & Yes & Yes  \\
WinRAR                 & 5.40                 & No  & No   \\
Oracle VM VirtualBox   & 4.0.16               & Yes & Yes  \\
Mozilla Firefox 19.0 beta1 & 19.0             & No  & Yes  \\
Skype 7.16             & 7.16.102            & No  & No   \\
\hline
\textbf{Total Detected} & -                    & 2 apps & 4 apps \\
\hline
\end{tabular}
\label{tab:application_sample_space}
\end{table}

\paragraph{Analysis of Sample Space 1} 
Table I highlights a sample of six applications tested for vulnerabilities. FleetDM detected vulnerabilities in only 2 applications, while Vajra (with improved sanitization and fuzzy matching) detected vulnerabilities in 4 out of the 6 applications. Key observations include:
\begin{itemize}
    \item \textbf{Adobe Acrobat (64-bit)}: Marked as vulnerable in Vajra due to enhanced sanitization processes that corrected inconsistencies in the vendor name.
    \item \textbf{VLC Media Player}: Initially detected by FleetDM but missed in Vajra’s previous version. The new Vajra version successfully identified the vulnerability after adding title sanitization.
    \item \textbf{Oracle VM VirtualBox}: Detected as vulnerable by both FleetDM and Vajra.
    \item \textbf{Mozilla Firefox 19.0 beta1}: Detected only in Vajra  due to fuzzy matching improvements.
\end{itemize}

These findings indicate that our enhanced Vajra system, particularly after recent updates, can detect vulnerabilities that FleetDM may overlook, particularly in applications with inconsistent names or non-standard versions.

\subsection{Detection Rate Comparison}
To evaluate the overall effectiveness of both systems, we conducted additional testing on a larger sample of 10 applications. The detection rates for FleetDM and Vajra were calculated based on the number of vulnerabilities identified in each system. Table~\ref{tab:sample_space2} presents a summary of the results.\begin{table}[H]
\centering
\caption{Application Sample Space 2 - Vulnerability Detection Summary}
\begin{tabular}{|c|c|c|c|}
\hline
\textbf{Application Name}       & \textbf{Version}              & \textbf{FleetDM}& \textbf{Vajra} \\
\hline
Winamp                          & 5.541                         & Yes            & Yes            \\
Foxit PDF Reader                & 5.4.5.0124                    & Yes            & Yes            \\
Notepad++                       & 5.9.6.2                       &                &                \\
Postman                         & v7.26.1 (64 bit Windows)      &                & Yes            \\
Teams                           & 24165.1306.2686.9504          &                &                \\
OpenVPN                         & 2.4.11-I062.win10            &                & Yes            \\
Webex Teams                     & v4.19.3.29764                 &                &                \\
iTunes                          & 11.0.1.12 (x64)               & Yes            & Yes            \\
VMware Server                   & 1.0.7                         & Yes            & Yes            \\
Macromedia Flash Player         & 8.0.22.0                      & Yes            & Yes            \\
\hline
\textbf{Total Detected}         & -                             & 5              & 7              \\
\hline
\end{tabular}
\label{tab:sample_space2}
\end{table}

\paragraph{Detection Rate Calculations}
The detection rates for each system were calculated using the total applications tested and vulnerabilities detected:
\begin{itemize}
    \item \textbf{FleetDM Detection Rate}: Calculated as 50\%, based on 5 vulnerabilities detected out of 10 applications tested.
    
    \item \textbf{Vajra Detection Rate}: Calculated as 70\%, based on 7 vulnerabilities detected out of 10 applications tested.
\end{itemize}

This comparison shows that Vajra’s detection rate was 20\% higher than that of FleetDM, underscoring the benefits of enhanced sanitization and fuzzy matching.

\subsection{Improvement Analysis}
The improvement in detection rates reflects the impact of Vajra's sanitization and fuzzy matching processes, which address inconsistencies that FleetDM's simpler approach may miss. We can quantify the improvement as follows:

\paragraph{Improvement Rate Calculation}
The improvement rate between FleetDM and Vajra’s detection rates is calculated as a 40\% increase over FleetDM's detection rate.

This indicates that Vajra’s advanced methodologies increased the detection accuracy by 40\% compared to FleetDM.

\paragraph{Impact of Sanitization and Fuzzy Matching}
The enhancements in detection accuracy are attributed to:
\begin{itemize}
    \item \textbf{Enhanced Sanitization Techniques}: By standardizing software names, vendor names, and versions, Vajra can accurately match applications that may have slightly different naming conventions in the NVD.
    \item \textbf{Fuzzy Matching Algorithm}: With RapidFuzz’s similarity scoring, Vajra tolerates minor variations in software names, reducing false negatives and ensuring that highly similar CPE strings are detected.
\end{itemize}

The combined effect of sanitization and fuzzy matching increased Vajra's detection rate from 50\% to 70\%, reflecting a 20\% increase in detection accuracy over FleetDM. This improvement is especially valuable in detecting vulnerabilities in legacy or less commonly updated software versions that may not have direct matches in traditional CPE databases.

\section{Limitations and Future Work}

This section outlines the current limitations of our methodology, particularly in cases that challenge vulnerability detection accuracy. It also proposes future enhancements aimed at improving the system's performance, adaptability, and coverage.

\subsection{Limitations}

While our approach has improved accuracy in CPE string generation and vulnerability detection, several limitations remain:

\paragraph{1. Handling of Non-Standard Application Names}
Despite sanitization and fuzzy matching, some applications with unique or non-standardized names fall below the threshold for accurate CPE generation. For instance, the program ``WinRAR 5.20 (32-bit)'' by ``win.rar GmbH'' is listed in the CPE dictionary as ``RARLAB WinRAR 5.20'' by ``rarlab,'' with the CPE string:

\begin{verbatim}
cpe:2.3:a:rarlab:winrar:5.20:*:*:*:*:*:*:*
\end{verbatim}

These discrepancies can hinder the matching process. Custom rules for adjusting program names and vendors during sanitization are required to ensure correct CPE generation, particularly for applications with unique identifiers.

\paragraph{2. Threshold Sensitivity in Fuzzy Matching}
Balancing the sensitivity of fuzzy matching thresholds is challenging. High thresholds may miss legitimate software matches (false negatives), while low thresholds can lead to incorrect matches (false positives). Fine-tuning thresholds based on software characteristics could improve the balance between sensitivity and specificity.

\paragraph{3. Dependency on Updated CPE Dictionaries}
The accuracy of our methodology depends on up-to-date CPE dictionaries. If the dictionary is outdated or lacks entries, vulnerability matching is compromised. Frequent updates are essential to maintain effectiveness in vulnerability detection.

\subsection{Future Work}

To address these limitations and enhance detection accuracy, efficiency, and scalability, we propose the following improvements:

\paragraph{Real-Time Data Integration}
Incorporating real-time data ingestion (e.g., via Apache Kafka) could ensure timely updates to the CPE dictionary and vulnerability data, allowing the system to respond more quickly to emerging threats and keep vulnerability information up-to-date.

\paragraph{Expanded Vulnerability Sources}
Currently, the system primarily relies on the National Vulnerability Database (NVD). Integrating additional sources, such as RPM and NPM repositories, would extend coverage to vulnerabilities in Linux and JavaScript ecosystems, making the detection framework more comprehensive.

\paragraph{Automated Rule-Based Pipeline for Custom Applications}
Developing an automated rule-based pipeline would streamline detection for applications with unique naming conventions. This pipeline would standardize entries (e.g., adjusting ``WinRAR'' by ``win.rar GmbH'' to match ``RARLAB WinRAR'' by ``rarlab'') for accurate CPE generation. The system could identify and apply necessary transformations for known mismatches automatically, enhancing detection accuracy without manual intervention.

\paragraph{Adaptive Fuzzy Matching Thresholds}
Implementing adaptive thresholds for fuzzy matching could improve detection accuracy by adjusting sensitivity based on software attributes. For widely used software with established naming conventions, stricter thresholds would reduce false positives, while more lenient thresholds could be applied to ambiguous entries, minimizing false negatives.

\section{Conclusion}
In this paper, we presented an enhanced methodology for software vulnerability detection by addressing the inconsistencies in Common Platform Enumeration (CPE) strings through sanitization and fuzzy matching. Our approach tackles the limitations of traditional open-source tools like FleetDM by normalizing software names, vendor information, and version numbers, thereby aligning non-standardized CPE strings with database entries more effectively.

Our solution demonstrated a 40\% improvement in detection accuracy over baseline methods, highlighting the importance of multi-step sanitization combined with similarity-based fuzzy matching. The union query approach, prioritized by relevance, enabled precise filtering, while fuzzy matching tolerated minor discrepancies, ensuring higher reliability in detecting vulnerabilities. 

This work underscores the impact of adaptive matching algorithms in cybersecurity, with future potential for real-time data ingestion, rule-based adjustments for unique software entries, and expansion to broader vulnerability sources. By integrating these advancements, the proposed methodology offers a scalable and robust framework for enhanced vulnerability management, equipping organizations to proactively mitigate security risks in increasingly diverse software environments.

\bibliographystyle{IEEEtran}
\bibliography{references}









\end{document}